\documentclass[twocolumn,aps,pra,superscriptaddress,showpacs,tightenlines]{revtex4-1}
\usepackage{setspace}
\usepackage{mathptmx}
\DeclareMathAlphabet{\mathcal}{OMS}{cmsy}{m}{n}
\usepackage{mathrsfs}

\usepackage[T1]{fontenc}
\usepackage[latin9]{inputenc}
\usepackage{amstext}
\usepackage{amsmath}
\usepackage{graphicx}
\usepackage{esint}
\usepackage{color}
\usepackage[colorlinks, linkcolor=blue, citecolor=blue,hyperindex,bookmarks=false,pdfstartview=FitH]{hyperref}
\usepackage{lineno}
\makeatletter

\@ifundefined{textcolor}{}
{%
 \definecolor{BLACK}{gray}{0}
\definecolor{WHITE}{gray}{1}
 \definecolor{RED}{rgb}{1,0,0}
 \definecolor{GREEN}{rgb}{0,1,0}
 \definecolor{BLUE}{rgb}{0,0,1}
 \definecolor{CYAN}{cmyk}{1,0,0,0}
 \definecolor{MAGENTA}{cmyk}{0,1,0,0}
 \definecolor{YELLOW}{cmyk}{0,0,1,0}
}
\makeatother

\begin{document}

\title{Robust Preparation of Many-body Ground States in Jaynes-Cummings Lattices}

\author{Kang Cai}
\affiliation{School of Natural Sciences, University of California, Merced, California 95343, USA}

\author{Prabin Parajuli}
\affiliation{School of Natural Sciences, University of California, Merced, California 95343, USA}

\author{Guilu Long}
\affiliation{Beijing Academy of Quantum Information Sciences, Beijing 100193, China}
\affiliation{State Key Laboratory of Low-Dimensional Quantum Physics and Department of Physics, Tsinghua University, Beijing 100084, China}
\affiliation{Beijing National Research Center for Information Science and Technology and School of Information, Tsinghua University, Beijing 100084, China}
\affiliation{Frontier Science Center for Quantum Information, Beijing 100084, China}

\author{Chee Wei Wong}
\affiliation{Electrical \& Computer Engineering, and Center for Quantum Science \& Engineering, University of California, Los Angeles, California 90095, USA}

\author{Lin Tian} 
\email[email:]{ltian@ucmerced.edu} 
\affiliation{School of Natural Sciences, University of California, Merced, California 95343, USA}

\begin{abstract}
Strongly-correlated polaritons in Jaynes-Cummings (JC) lattices can exhibit quantum phase transitions between the Mott-insulating and superfluid phases at integer fillings. The prerequisite to observe such phase transitions is to pump polariton excitations into a JC lattice and prepare them into appropriate ground states. Despite previous efforts, it is still challenging to generate many-body states with high accuracy. 
Here we present an approach for the robust preparation of many-body ground states of polaritons in finite-sized JC lattices by optimized nonlinear ramping. We apply a Landau-Zener type of estimation to this finite-sized system and derive the optimal ramping index for selected ramping trajectories, which can greatly improve the fidelity of the prepared states. With numerical simulation, we show that by choosing an appropriate ramping trajectory, the fidelity in this approach can remain close to unity in almost the entire parameter space. This approach can shed light on high-fidelity state preparation in quantum simulators and advance the implementation of quantum simulation with practical devices.
\end{abstract}


\maketitle

{\parindent 0 pt \bf INTRODUCTION} 
\vskip 2mm

{\parindent 0 pt The Jaynes-Cummings (JC) model is a prototype for studying light-matter interaction, where a quantum two-level system is coupled to a cavity mode~\cite{JCmodel}. This model has been utilized to study cavity or circuit quantum electrodynamics (QED) in a wide range of systems, from individual particles in the atomic scale to collective modes in mesoscopic devices~\cite{cavityQED, circuitQED1, circuitQED2}. More recently, advances in device fabrication and quantum technology enabled the exploration of many-body physics in arrays of JC models, i.e., JC lattices, which can be realized with optical cavities coupled to defects in semiconductors~\cite{Hartmann:2006, Greentree:2006, Angelakis:2007, 2007RossiniPRL_JC, 2008NeilPra_BH} and superconducting circuit QED systems~\cite{2009KochPra_QS, 2012HouckNP_JCQS, TianPRL2011, Seo2015:1, 2015TianScienceChina_QS, Xue2017}. The light-matter coupling in a JC model induces intrinsic nonlinearity in the energy spectrum, which can be mapped to an onsite repulsive interaction between polariton excitations. The competition between this onsite interaction and polariton hopping between neighboring sites gives rise to rich many-body physics for strongly-correlated polaritons in JC lattices, such as quantum or dissipative phase transitions and photon blockade effects~\cite{Hoffman:2011, KeelingPRL2012, HouckPRX2017}. Moreover, when the counter-rotating terms in the qubit-cavity interaction cannot be neglected, the system becomes a quantum Rabi lattice, where distinctively different many-body phase transitions have been studied~\cite{TureciPRL2012, JalalPRA2013}.
One effect of particular interest is the quantum phase transitions between the Mott-insulating (MI) and superfluid (SF) phases for polaritons in JC lattices at integer fillings, featured by the occurrence of off-diagonal long-range order in the correlation functions. It was shown that such phase transitions can be observed in coupled cavity arrays~\cite{Hartmann:2006, Greentree:2006, Angelakis:2007} and multi-connected JC lattices~\cite{Seo2015:1, 2015TianScienceChina_QS, Xue2017}. }

The prerequisite to observe the MI-SF phase transitions is to pump polariton excitations into a JC lattice and prepare them into appropriate ground states. However, preparing many-body ground states is a challenging task in engineered systems such as quantum simulators~\cite{Feynman, Lloyd, AspuruGuzik2005} and adiabatic quantum computers~\cite{Farhi2000_1, Albash2016}.  A number of approaches have been studied to tackle this problem, including adiabatic quantum evolution~\cite{FarhiScience2001, RolandCerfPRA2002, HTQuanNJP2010}, quantum shortcut method by applying  counter-diabatic interactions~\cite{XChenPRL2010, delCampoPRL2012}, quantum phase estimation via quantum Fourier transformation~\cite{Kitaev1995, Abrams1997}, variational quantum eigensolver~\cite{Peruzzo2014, Dumitrescu2018}, full quantum eigensolver~\cite{LongResearch2020}, and engineered dissipative environment for the preparation and stabilization of entangled states~\cite{Kraus2008, Verstraete2009, TureciPRA2014, TureciPRX2016}. Despite these efforts, it is still hard to generate desired many-body states with high fidelity in the noisy intermediate-scale quantum (NISQ) era~\cite{Preskill}, in particular, for systems working with excitations such as the JC lattices. The barriers to generating desired many-body states efficiently and accurately include the lack of a priori knowledge of the energy spectrum, the difficulty in engineering complicated counter-diabatic interactions, the rapid decrease of the energy gap and quick increase of the dimension of the Hilbert space with the size of the quantum simulators, and the finite decoherence times in NISQ devices. Furthermore, many-body states in strongly-correlated systems can be highly entangled, unknown, and hence, often impossible to be generated with pre-programed quantum logic gates. 

Here we study the robust generation of many-body ground states in finite-sized JC lattices at unit filling using optimized nonlinear ramping. 
In previous works~\cite{SenPRL2008, MondalPRB2009, BarankovPRL2008}, it was shown that nonlinear ramping can reduce diabatic transitions to excited states or the production of domain walls when a many-body system in the thermodynamic limit evolves across a quantum critical point due to the scaling of the phase transition. We apply nonlinear ramping to a finite-sized system, where the energy gap between the ground and the excited states remains finite. By exploiting a Landau-Zener type of estimation~\cite{lz1, lz2} and the spectral feature along a selected ramping trajectory, we derive the optimal ramping index for the trajectory, which can significantly improve the fidelity of the prepared state. Our estimation agrees well with the result from our numerical simulation of the ramping process. Moreover, we show that by selecting an appropriate trajectory for a given set of target parameters in combination with the optimal ramping index, the fidelity can remain close to unity in almost the entire parameter space. The ramping trajectory can be adjusted by varying the initial parameters or the ratios between the ramping indices for different parameters.
The initial states of this nonlinear ramping process can be prepared with high accuracy by applying engineered pulse sequences~\cite{1996EberlyPRL_StatePrepare} when tuning the system parameters to either the deep MI regime with no hopping between adjacent unit cells or the deep SF regime with diminishing light-matter coupling. 

JC lattices have been implemented with superconducting quantum devices in recent experiments~\cite{HouckPRX2017, NeillScience2018, YYu2018}, and the ramping process studied here is within reach of current technology~\cite{squbit_rev1, squbit_rev2, squbit_rev3}. Using practical parameters from the experiments, we show that high fidelity can be achieved for the prepared states on a time scale much shorter than the observed decoherence times of these devices. Meanwhile, the approach of optimized nonlinear ramping for finite-sized systems is general and can be applied to many other models. The study of state generation in finite-sized systems with this approach can provide insights to the problem of preparing complex quantum states in quantum computers. Our result can hence shed light on the high-fidelity preparation of many-body states in engineered quantum systems such as quantum simulators, and advance the implementation of quantum simulation with NISQ devices. 
\vskip 4mm

{\parindent 0 pt \bf RESULTS} 
\vskip 2mm

{\parindent 0 pt \bf Model and quantum phase transition}

{\parindent 0 pt Consider the JC lattice depicted in Fig.~1a. Here each unit cell contains a qubit coupled to a cavity mode with coupling strength $g$, and adjacent unit cells are connected via photon hopping with hopping rate $J$. The total Hamiltonian of this model is $H_{\rm t}=H_{0}+H_{\rm int}$ ($\hbar=1$), where
\begin{equation}
H_{0} = \omega _{\rm c}\sum_{j}a_{j}^{\dagger }a_{j}+\omega _{z}\sum_{j}\frac{\sigma _{jz}+1}{2} + g \sum_{j}\left(a_{j}^{\dagger }\sigma _{j-}+\sigma _{j+}a_{j} \right)\label{eq:H0}
\end{equation}
is the Hamiltonian of uncoupled JC models with $\omega_{\rm c}$ the frequency of cavity modes, $\omega _{z}$ the level splitting of the qubits, $a_{j}$ ($a_{j}^{\dagger}$) the annihilation (creation) operator of the $j$-th cavity mode, and $\sigma_{j\pm},\, \sigma_{jz}$ the Pauli operators of the $j$-th qubit, and 
\begin{equation}
H_{\rm int} =-J \sum_{j}\left( a_{j}^{\dagger}a_{j+1}+a_{j+1}^{\dag}a_{j}\right)\label{eq:Hint}
\end{equation}
is the photon hopping between neighboring unit cells. 
\begin{figure}[t]
\centering
\includegraphics[width=8.5cm, clip]{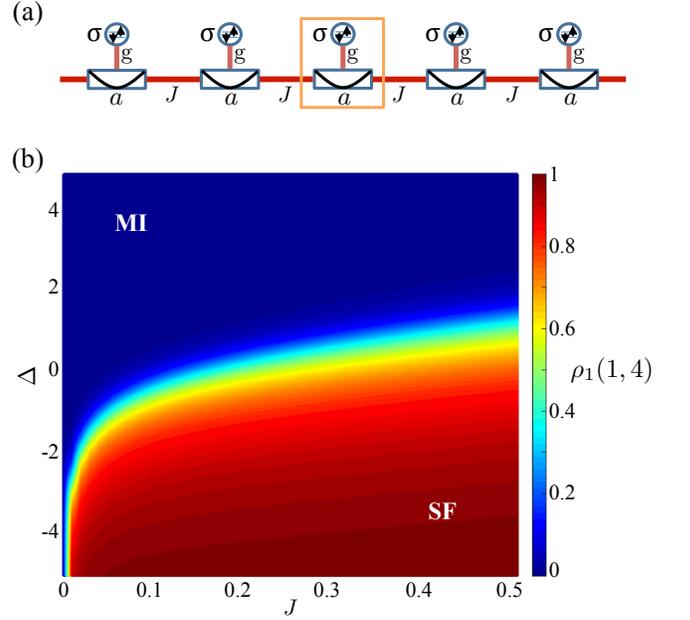}
\caption{{\bf Quantum phase transition in JC lattice. a} Schematic of a 1D JC lattice. Circles (rectangles) represent qubits (cavity modes) with light-matter coupling $g$ and hopping rate $J$. {\bf b} Single-particle density matrix $\rho_{1}(1,4)$ vs hopping rate $J$ and detuning $\Delta$ for a finite-sized lattice at unit filling with $N=L=6$. Here we let $g$ be the energy unit with $g\equiv1$.}
\label{fig1}
\end{figure}
Let $\{\vert n, s\rangle\}$ be the basis set of an individual JC model with the cavity in the Fock state of photon number $n$ and the qubit in the state $s=\uparrow, \downarrow$. The eigenstates of the JC model include the ground state $\vert g_{0}\rangle =\vert 0, \downarrow\rangle$ with no excitation and the polariton doublets $\vert n,\pm\rangle$ with $n$ excitations:
\begin{eqnarray}
|n,+\rangle &=& \cos\left(\theta/2\right)|n,\downarrow\rangle + \sin\left(\theta/2\right)|n-1,\uparrow\rangle, \label{eq:np} \\
|n,-\rangle &=&  \sin\left(\theta/2\right)|n,\downarrow\rangle - \cos\left(\theta/2\right)|n-1,\uparrow\rangle, \label{eq:nm} 
\end{eqnarray}
where $\theta=2\arcsin{\sqrt{\left[1-\Delta /\chi(n)\right]/2}}$, $\chi(n)=\sqrt{\Delta^{2}+4ng^{2}}$, and $\Delta =\omega _{\rm c}-\omega_{z}$ is the detuning between the cavity mode and the qubit. The corresponding eigenenergies are $E_{g_{0}} =0$ and $E_{n,\pm} =\left( n-1/2\right) \Delta \pm \chi(n)/2$. When the coupling $g$ is nonzero, the energy spacings between these eigenstates are unequal with $(E_{n+1,-}-E_{n,-})>(E_{n,-}-E_{n-1,-})$. Specifically, $E_{2,-}>2E_{1,-}$ for $n=1$, which indicates that the energy to add two polaritons to the JC model is more than twice the energy to add a single polariton. The extra energy to add a second polariton can be viewed as an effective onsite interaction or nonlinearity for the polaritons, which is at the root of many interesting phenomena in JC models or JC lattices, such as the photon blockade effect~\cite{Hartmann:2006, Greentree:2006, Angelakis:2007} and electron-phonon-like effects~\cite{TianPRB2013}.}

In the limit of $J=0$, the JC lattice is composed of isolated JC models. The ground state at unit filling, where the number of polaritons $N$ is equal to the number of lattice sites $L$, is
\begin{equation}
|G\rangle_{J=0}=\prod_{j} \vert 1,-\rangle _{j}\label{eq:GJ0}
\end{equation}
with one polariton excitation occupying the state $\vert 1,-\rangle$ per site, which is in the deep MI regime. States with more than one excitation at the same site are energetically unfavorable due to the effective onsite interaction. In the opposite limit of $g=0$ at finite hopping rate $J$, the cavity modes are decoupled from the qubits. The hopping Hamiltonian (\ref{eq:Hint}), now the dominant term, can be transformed to the momentum space under the periodic boundary condition with $H_{\rm int}=-2 J \sum_{k}\cos \left(k\right) a_{k}^{\dagger }a_{k}$, where $a_{k}$ ($a_{k}^{\dagger}$) is the annihilation (creation) operator of a collective cavity mode at the quasimomentum $k ={\rm \pi} m /N$ with integer $m\in \left [-(N-1), N\right]$ and $a_{k}=\sum_{j}a_{j}{\rm e}^{{\rm i} k\cdot j} /\sqrt{N}$. At $\Delta < 0$ with the cavity frequency below the qubit energy splitting, the ground state at unit filling is
\begin{equation}
|G\rangle _{g=0} =\frac{1}{\sqrt{N!}}\left( a_{k=0}^{\dagger }\right)^{N}\prod_{j}\vert0,\downarrow\rangle_{j}\label{eq:Gg0}
\end{equation}
with all polaritons occupying the $k=0$ mode, which is a nonlocal state in the deep SF regime. 

With the mean-field approximation~\cite{Hartmann:2006, Greentree:2006, Angelakis:2007} and numerical methods~\cite{2007RossiniPRL_JC, 2008NeilPra_BH, Seo2015:1, 2015TianScienceChina_QS, Xue2017}, it was shown that quantum phase transitions between the MI and SF phases due to the competition between the onsite interaction and the photon hopping can occur in the intermediate regimes of the parameter space in JC lattices. For a finite-sized lattice with $N=L=6$, we numerically calculate the many-body ground states using the exact diagonalization method. In this finite-sized system, the energy separation between the ground and the excited states decreases as the parameters approach the intermediate regimes, but maintains a finite energy gap. The spatial correlation in the many-body ground state $\vert G\rangle$ can be characterized by the normalized single-particle density matrix defined as~\cite{Penrose:1956, Yang:1962}
 \begin{equation}
\rho_{1}(i,j)=\langle G|a^{\dagger}_{i}a_{j}|G\rangle/\langle G|a^{\dagger}_{i}a_{i}|G\rangle,\label{eq:rho1}
\end{equation}
which reveals the off-diagonal long-range order of the state. The single-particle density matrix decreases algebraically with the spatial separation $\vert i-j\vert$ in the SF phase and decreases exponentially in the MI phase~\cite{Seo2015:1, 2015TianScienceChina_QS, Xue2017}. For a finite $\vert i-j\vert$, $\rho_{1}(i,j)$ of the SF phase is much larger than that of the MI phase. In Fig.~1b, we plot our numerical result of $\rho_{1}(1,4)$ as functions of the hopping rate $J$ and the detuning $\Delta$, with the coupling $g$ as the energy unit ($g\equiv1$). It can be seen that $\rho_{1}(1,4)$ increases with $J$ at arbitrary detuning. In the deep MI regime with $J=0$, $\rho_{1}(1,4)=0$ with the polaritons localized in the lattice. In the deep SF regime, $\rho_{1}(1,4)$ can approach unity. This result clearly indicates the occurrence of the MI-SF phase transition in the thermodynamic limit. 
\vskip 2mm

{\parindent 0 pt \bf State initialization}

{\parindent 0 pt We present methods to pump $N=L$ polaritons to the JC lattice in the limiting cases of $J=0$ and $g=0$, respectively, by applying engineered pulses. The polaritons are pumped into the many-body ground states at the corresponding parameters. These states will be used as the initial state of the nonlinear ramping approach.}

In the deep MI limit of $J=0$ and finite $g$, the ground state is given by (\ref{eq:GJ0}) with each unit cell in the polariton state $\vert 1,-\rangle$. Because the unit cells are decoupled, we can perform a Rabi rotation between the states $\vert g_{0}\rangle$ and $\vert 1,-\rangle$ on each JC model, as illustrated in Fig.~2a. The driving Hamiltonian can have the form $H_{\rm d1}\left( t\right) = \sum_{j}  [ \varepsilon{\rm e}^{{\rm i}\omega_{\rm L}t}\sigma_{j-}+h.c.]$ with driving amplitude $\varepsilon$ and driving frequency $\omega _{\rm L}=E_{1,-}-E_{g_{0}}$. The corresponding Rabi frequency can be derived as $\Omega_{\rm d1}= \vert\varepsilon\cos(\theta/2)\vert$ following Eq.~(\ref{eq:nm}). The duration of the Rabi flip from the initial state $\vert g_{0}\rangle$ to the final state $\vert 1,-\rangle$ is $\tau_{\rm d1}={\rm \pi}/2\Omega_{\rm d1}$. To prevent the driving pulse from inducing unwanted transitions to higher states such as $ \vert 1,+\rangle $, it requires that $\vert \varepsilon\vert \ll g$.

\begin{figure}[t]
\begin{center}
\includegraphics[width=8.3cm,clip]{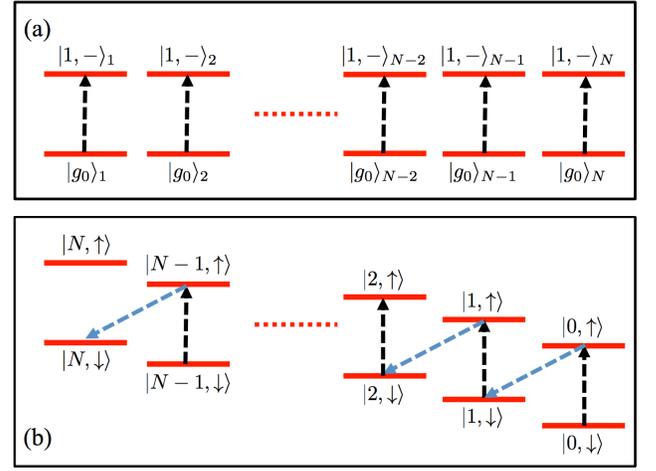}
\caption{{\bf Pulse sequence for state initialization. a} Pulses for MI initial state at $J=0$ and finite $g$. The vertical arrows are Rabi flips between the states $\vert g_{0}\rangle$ and $\vert 1,-\rangle$ in each JC model. {\bf b} Pulses for SF initial state at $g=0$ and finite $J$. The vertical (slanted) arrows are the operations $C_{l}$ ($Q_{l}$) with $l\in [1, N]$ on the coupled system of the auxiliary qubit and mode $a_{k=0}$.}
\label{fig2}
\end{center}
\end{figure}
In the deep SF limit of $g=0$ and finite $J$, the ground state is given by (\ref{eq:Gg0}) with all polaritons occupying the collective (nonlocal) mode $a_{k=0}$. To generate this state, we introduce an auxiliary qubit with Pauli operators $\sigma_{0\pm}$, $\sigma_{0z}$. This qubit has the Hamiltonian 
\begin{equation}
H_{\rm d2}\left(t\right) = \frac{\omega_{0}}{2}\sigma_{0z}+\varepsilon\left( t\right) {\rm e}^{{\rm i}\omega_{\rm L}t}\sigma _{0-}+g_{\rm d}\left( t\right) \sum_{j}a_{j}^{\dag}\sigma _{0-}+h.c.,\label{eq:Hd2}
\end{equation}
which includes the qubit energy splitting $\omega_{0}$, a driving on the qubit with amplitude $\varepsilon\left( t\right)$ and frequency $\omega_{\rm L}$, and a tunable coupling between the qubit and the cavity modes with coupling strength $g_{\rm d}\left( t\right)$. By choosing $\omega_{\rm L},\,\omega_{0}$ to be both in resonance with the mode $a_{k=0}$, we have $H_{\rm d2, r}\left( t\right) =\varepsilon\left( t\right)\sigma _{0-} + \sqrt{N}g_{\rm d}\left( t\right)a_{k=0}^{\dagger }\sigma _{0-} +h.c.$ in the rotating frame. The first term of $H_{\rm d2, r}$ generates a Rabi rotation on the auxiliary qubit, and the second term is the coupling between the auxiliary qubit and the mode $a_{k=0}$. Both terms can be turned on and off within nanoseconds, as has been demonstrated in recent experiments on superconducting transmon qubits~\cite{NeillScience2018, YYu2018}.
As the qubits in the JC lattice are decoupled from the cavities, the state of mode $a_{k=0}$ is only affected by its coupling to the auxiliary qubit. Let the initial state of the coupled system of mode $a_{k=0}$ and the auxiliary qubit be $\vert 0, \downarrow\rangle$. To generate the state (\ref{eq:Gg0}), we utilize the approach in \cite{1996EberlyPRL_StatePrepare} to design a pulse sequence, which can be implemented by switching on $\varepsilon\left( t\right)$ and $g_{\rm d}\left( t\right)$ alternately. The unitary operator for this pulse sequence is
\begin{equation}
U =Q_{N}C_{N}Q_{N-1}C_{N-1}\cdots Q_{2}C_{2}Q_{1}C_{1},\label{eq:U}
\end{equation}
where the unitary operator $C_{l}$ ($l\in [1, N]$) incurs a Rabi flip on the auxiliary qubit by applying a driving pulse with amplitude $\varepsilon$ for a duration of $\tau_{\rm cl}={\rm \pi}/2\vert \varepsilon\vert$, and the unitary operator $Q_{l}$ enables the exchange of excitations between the auxiliary qubit and the mode $a_{k=0}$ by turning on the coupling $g_{\rm d}$ for a duration of $\tau_{\rm ql}={\rm \pi}/2\sqrt{N l}\vert g_{\rm d}\vert$. Following this pulse sequence, the state evolves as $\vert 0, \downarrow\rangle \rightarrow \vert 0, \uparrow\rangle \rightarrow \vert 1, \downarrow\rangle \cdots \rightarrow\vert N, \downarrow\rangle$, as shown in Fig.~2b. 
The total duration of this pulse sequence is $\tau_{\rm d2}=\sum_{l}\left(\tau_{\rm cl}+\tau_{\rm ql}\right)$. Assuming that the magnitudes of $\varepsilon$ and $g_{\rm d}$ are the fixed for all $l$'s, we find $\tau_{\rm d2}=N{\rm \pi}/2\vert \varepsilon\vert + \sum_{l}{\rm \pi}/2\sqrt{Nl}\vert g_{\rm d}\vert$, which increases with the total number of polaritons as $\tau_{\rm d2}={\rm O}(N)$. Meanwhile, it requires that $\vert \varepsilon\vert, \sqrt{N}\vert g_{\rm d} \vert \ll \omega_{\rm L}$ to achieve high fidelity for the generated state. Note that the other collective modes $a_{k\ne0}$ of the cavities are not coupled to the auxiliary qubit due to the symmetry of the Hamiltonian $H_{\rm d2}$. The excitation of these modes will not occur during this pulse sequence. 
\vskip 2mm
 
{\parindent 0 pt \bf Optimized nonlinear ramping}

{\parindent 0 pt Many-body ground states in the intermediate regimes of the parameter space cannot be calculated analytically, and we cannot design quantum logic operations to generate such states, in contrast to the ground states in the deep MI or SF regimes. We employ optimized nonlinear ramping to reach such states via adiabatic evolution. 
In this approach, a parameter $p$ has the time dependence:
\begin{equation}
p(t) = p(0)\left[1-\left(t/T\right)^{r_{p}}\right] + p(T)\left(t/T\right)^{r_{p}}, \label{eq:parat}
\end{equation}
where $p=g, J, \Delta$ is a tunable parameter of the JC lattice, $p(0)$ is the initial value of the parameter at time $t=0$, $p(T)$ is the target value at the final time $T$, and $r_{p}$ is the ramping index of parameter $p$. For $r_{p}=1$, it is the linear ramping studied in \cite{Farhi2000_1, Albash2016}; and $r_{p}\ne 1$ corresponds to nonlinear ramping~\cite{SenPRL2008, MondalPRB2009, BarankovPRL2008}. It can be shown that for any parameter $p$ at an arbitrary time $t$, 
\begin{equation}
\left(\frac{p(t)- p(0)}{p(T)- p(0)}\right)^{1/r_{p}} \equiv \left(\frac{J(t)-J(0)}{J(T)- J(0)}\right)^{1/r_{J}}. \label{eq:path}
\end{equation}
Hence when the initial and target parameters are given, the ramping trajectory in the parameter space is only determined by the ratios of the ramping indices $r_{g}/r_{J}$ and $r_{\Delta}/r_{J}$ and is independent of the specific value of an individual ramping index (see Supplementary Notes).
On the other hand, the value of an individual ramping index can strongly affect the sweeping rate of the Hamiltonian along a given trajectory. 
The sweeping rate of the Hamiltonian at parameters $\{p\}$ can be written as $\langle dH/dt \rangle = \sum_{p}\langle \partial H/\partial p \rangle p^{\prime}(p)$, where $p^{\prime}(p)$ is the time derivative $p^{\prime}=dp(t)/dt$ at $p=p(t)$, with $\langle .\rangle$ denoting the operator average at the ground state of parameters $\{p\}$. Using (\ref{eq:parat}), we obtain: 
\begin{equation}
p^{\prime}(p) = \frac{r_{p}\left[ p-p(0)\right]^{(r_{p}-1)/r_{p}}}{T\left[p(T)-p(0)\right]^{-1/r_{p}}}.\label{eq:dJdtinp} 
\end{equation}
By varying the ramping index $r_{p}$, $p^{\prime}(p)$, and hence $\langle dH/dt \rangle$, can be tuned in a large range. 
We also find that $p^{\prime}(p)=J^{\prime}(J)(r_{p}/r_{J})[p-p(0)]/[J-J(0)]$, which reveals that $p^{\prime}(p)/J^{\prime}(J)$ at a given position only depends on the trajectory, i.e., it only depends on the ratios $r_{g}/r_{J}$ and $r_{\Delta}/r_{J}$ that define the trajectory.}

Let $|\psi(T)\rangle$ be the wave function of the final state of the evolution at time $T$. The fidelity of the final state can be defined as $F=|\langle \psi(T)|G_{T}\rangle|^{2}$ with $|G_{T}\rangle$ the many-body ground state at the target parameters $\{p(T)\}=\{g(T), J(T), \Delta(T)\}$. During a continuous evolution, the probability of diabatic transitions can be approximated by the Landau-Zener formula $\sim {\rm e}^{-{\rm \pi} E_{\rm gp}^{2}/2H_{\rm gp}^{\prime}}$~\cite{lz1, lz2}, where the energy gap $E_{\rm gp}$ is defined as the minimal energy separation between the ground and the excited states along the evolution trajectory, and $H_{\rm gp}^{\prime}=\langle dH/dt \rangle_{\rm gp}$ denotes the sweeping rate of the Hamiltonian at the position of the energy gap. To reach the desired state with high fidelity, the adiabatic criterion, commonly expressed as $H_{\rm gp}^{\prime}\ll E_{\rm gp}^{2}$, needs to be satisfied so that diabatic transitions are negligible. For a given trajectory, we can optimize the ramping indices $r_{p}$ to minimize $H_{\rm gp}^{\prime}$ so as to suppress diabatic transitions in the most vulnerable region of the evolution and improve the fidelity of the final state. With the relation between different $p^{\prime}(p)$'s, we find that $H_{\rm gp}^{\prime} = c_{p} p^{\prime}(p_{\rm gp})$ for any parameter $p$, with $p_{\rm gp}$ the value of the parameter $p$ at the position of the energy gap. Here $c_{p}$ only depends on the ratios $r_{g}/r_{J}$ and $r_{\Delta}/r_{J}$, not the individual ramping indices; whereas $p^{\prime}(p_{\rm gp})$ depends on the ramping index $r_{p}$ as shown in (\ref{eq:dJdtinp}). At the optimal ramping index $r_{p}^{\rm (min)}$, $\partial H_{\rm gp}^{\prime}/\partial r_{p}=0$. This leads to 
\begin{equation}
r_{p}^{\rm (min)} = \log{\left[\frac{p(T)-p(0)}{p_{\rm gp}-p(0)}\right]},\label{eq:rpmin}
\end{equation}
which only depends on the position of the energy gap for a given trajectory.
\vskip 2mm

{\parindent 0 pt \bf Numerical simulation}

\begin{figure}[t]
\includegraphics[width=8.5cm, clip]{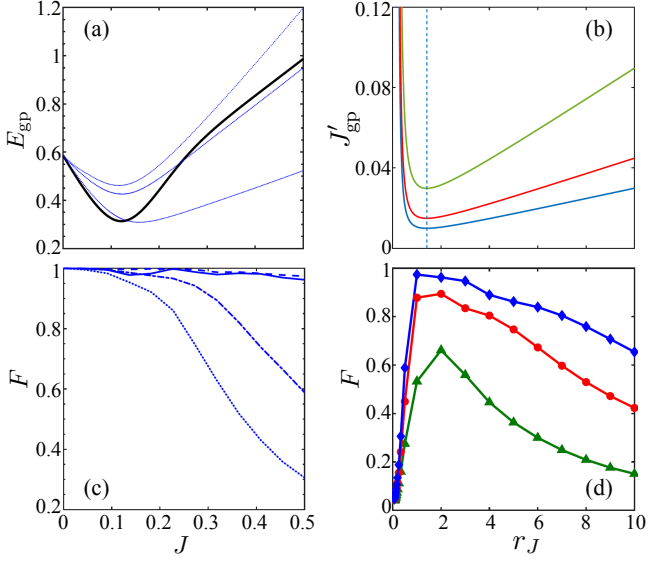}
\caption{{\bf Ramping from MI to SF phase. a} Energy spectrum of the lowest excited states vs hopping rate $J$. Solid (dotted) curve is for the symmetric (asymmetric) state with the ground-state energy set to zero. 
{\bf b} Time derivative $J^{\prime}_{\rm gp}$ vs ramping index $r_{J}$ at $J(T)=0.5$ and $T=5{\rm \pi}/g, 10{\rm \pi}/g, 15{\rm \pi}/g$ from top to bottom.
{\bf c} Fidelity $F$ vs $J(T)=J$ for $r_{J}=2$ (solid), $1$ (dashed), $1/2$ (dot-dashed), and $1/3$ (dotted) at $T=15{\rm \pi}/g$.
{\bf d} Fidelity $F$ vs $r_{J}$ at $J(T)=0.5$ and $T=5{\rm \pi}/g, 10{\rm \pi}/g, 15{\rm \pi}/g$ from bottom to top.
In all plots, $g(t)\equiv1$, $J(0)=0$, and $\Delta(t)\equiv0$.}
\label{fig3}
\end{figure}
{\parindent 0 pt Below we conduct numerical simulation to calculate the fidelity of the final states via nonlinear ramping with selected trajectories and compare the numerical result with the above estimation. In the simulation, we employ an algorithm developed in our previous work~\cite{Seo2015:1}, which only involves basis states with the total excitation number $N=L$. This algorithm greatly speeds up the calculation of the eigenstates and dynamics of JC lattices.
We first consider a trajectory following (\ref{eq:parat}) with $g(t)\equiv1$, $J(0)=0$, $J(T)=0.5$, and $\Delta(t)\equiv0$, where the photon hopping rate is continuously increased from zero to a finite value. The initial state is the deep MI phase in (\ref{eq:GJ0}). Using the exact diagonalization method, we can calculate the eigenstates and eigenenergies of the JC lattice along this trajectory. The energy spectrum of several lowest excited states is plotted as a function of the hopping rate $J$ in Fig.~3a. The solid curve corresponds to the energy of the lowest state that is symmetric with regard to all lattice sites, and the dotted curves are for asymmetric states. As both the initial state and the Hamiltonian $H(t)$ are symmetric with regard to lattice sites, the wave function $\vert \psi(t)\rangle$ at any time $t$ during the evolution must remain symmetric. Hence diabatic transitions can only happen between the ground state and symmetric states, and the energy gap related to the adiabatic criterion is determined by the energy separation between the ground state and the lowest symmetric state. From our numerical result, we find that the gap position is at $J_{\rm gp}=0.122$ with the energy gap $E_{\rm gp}=0.31$.}

The sweeping rate of the Hamiltonian can be written as $H_{\rm gp}^{\prime} = c_{J} J^{\prime}_{\rm gp}$ with $J^{\prime}_{\rm gp}$ the time derivative of the hopping rate $J$ at the gap position. Using (\ref{eq:dJdtinp}), we plot $J^{\prime}_{\rm gp}$ as a function of $r_{J}$ in Fig.~3b, where $J^{\prime}_{\rm gp}$ has a local minimum at the optimal ramping index $r_{J}^{\rm (min)}=\log\left[J(T)/J_{\rm gp}\right]$. For the selected trajectory, $r_{J}^{\rm (min)}=1.41$, which indicates that the best fidelity for the final state can be achieved with a ramping index in-between the linear and the quadratic forms. At a total evolution time $T=15{\rm \pi}/g$, the optimal ramping index gives $J^{\prime}_{\rm gp}=0.01$. With $E_{\rm gp}=0.31$, the adiabatic criterion is well satisfied. We numerically simulate this ramping process and calculate the fidelity of the final state. In Fig.~3c, the fidelity vs $J(T)=J$ is plotted for several values of $r_{J}$ at $T=15{\rm \pi}/g$. The fidelity decreases quickly with $J(T)$, as $J^{\prime}_{\rm gp}$ increases with $J(T)$. It can also be seen that for $J(T)$ sufficiently far away from $J_{\rm gp}$, where the Landau-Zener estimation becomes valid, the fidelity is much higher for $r_{J}=1,\, 2$ than that for $r_{J}=1/3,\,1/2$. As shown in Fig.~3d, the best fidelity for $J(T)=0.5$ is achieved when $r_{J}\in (1,\,2)$. These results agree well with our derivation of the optimal ramping index.

\begin{figure}[t]
\includegraphics[width=8.5cm, clip]{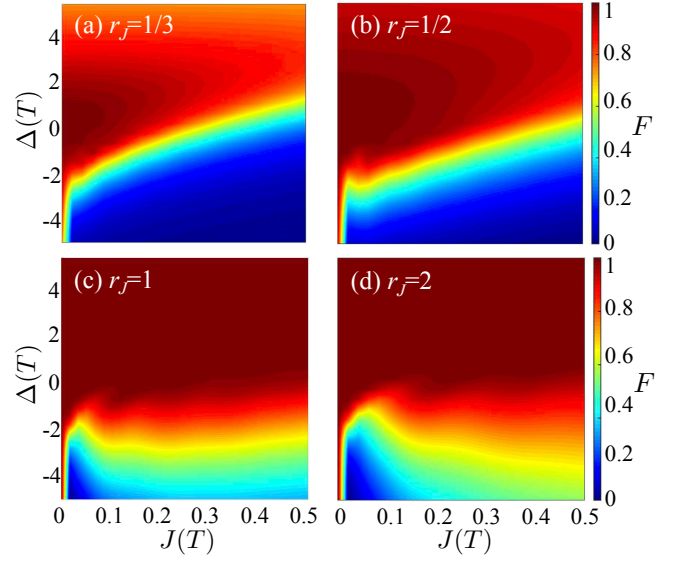}
\caption{{\bf Fidelity with MI initial state. a-d} Fidelity of the prepared state $F$ vs target hopping rate $J(T)$ and target detuning $\Delta(T)$ for ramping index $r_{J}$ given in the panel. Here $g(t)\equiv 1$, $J(0)=\Delta(0)=0$, $r_{\Delta}/r_{J}=1$, and $T=15{\rm \pi}/g$.}
\label{fig4}
\end{figure} 
We also obtain the fidelity of the final states for a wide range of target parameters $J(T),\,\Delta(T)$ following the trajectory (\ref{eq:parat}) with $g(t)\equiv1$, $J(0)=\Delta(0)=0$, $r_{\Delta}/r_{J}=1$, and the MI initial state (\ref{eq:GJ0}). The fidelity is presented in Fig.~4a-d for $r_{J}=1/3,\,1/2,\,1,\,2$, respectively. It can be seen that the fidelity decreases as the target parameters move further towards the SF phase. In particular, the fidelity exhibits a sharp decrease when the parameters cross the gap positions into the SF phase. Meanwhile, the fidelity demonstrates strong dependence on the ramping index in the intermediate regimes of the parameter space, which also agrees with our analytical prediction.
 
\begin{figure}[t]
\centering
\includegraphics[width=8.5cm, clip]{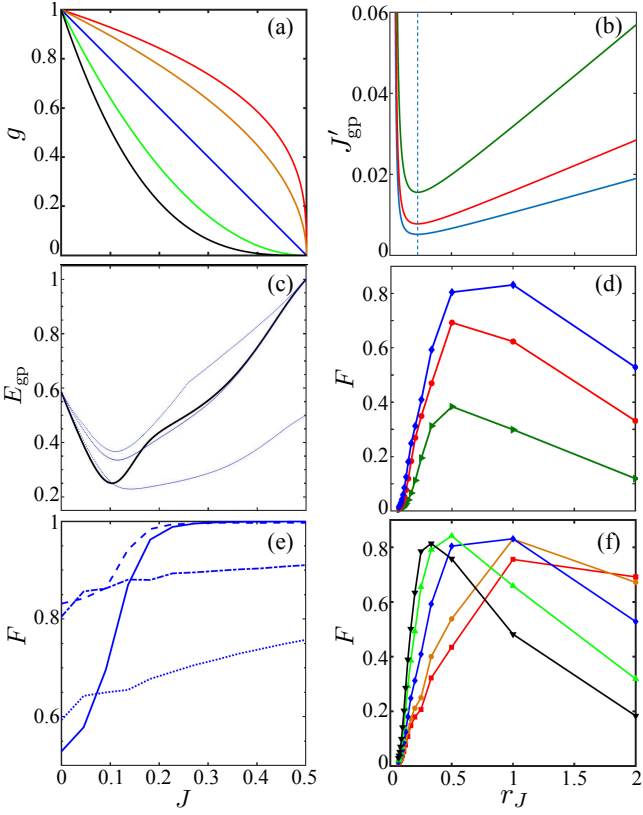}
\caption{{\bf Ramping from SF to MI phase. a} Ramping trajectory in the parameter space of $J$ and $g$ for $r_{g}/r_{J}=1/3, 1/2, 1, 2, 3$ from top to bottom.
{\bf b} Time derivative $J^{\prime}_{\rm gp}$ vs ramping index $r_{J}$ at $r_{g}/r_{J}=1$, $J(T)=0$, and $T=5{\rm \pi}/g, 10{\rm \pi}/g, 15{\rm \pi}/g$ from top to bottom.
{\bf c} Energy spectrum of the lowest excited states vs hopping rate $J$ for $r_{g}/r_{J}=1$. Solid (dotted) curve is for the symmetric (asymmetric) state with the ground-state energy set to zero. 
{\bf d} Fidelity $F$ vs $r_{J}$ at $r_{g}/r_{J}=1$, $J(T)=0$, and $T=5{\rm \pi}/g, 10{\rm \pi}/g, 15{\rm \pi}/g$ from bottom to top.
{\bf e} Fidelity $F$ vs $J(T)=J$ for $r_{J}=2$ (solid), $1$ (dashed), $1/2$ (dot-dashed), and $1/3$ (dotted) at $r_{g}/r_{J}=1$ and $T=15{\rm \pi}/g$.
{\bf f} Fidelity $F$ vs $r_{J}$ at $r_{g}/r_{J}=1/3$ (square), $1/2$ (circle), $1$ (diamond), $2$ (triangle), and $3$ (inverted triangle), $J(T)=0$, and $T=15{\rm \pi}/g$. 
In all plots, $g(0)=0$, $g(T)=1$, $J(0)=0.5$, and $\Delta(t)\equiv0$.}
\label{fig5}
\end{figure}
 Next we consider trajectories that starts from the deep SF phase with $g(0)=0$, $g(T)=1$, $J(0)=0.5$, $J(T)=0$, $\Delta(t)\equiv0$, and the initial state (\ref{eq:Gg0}). With both $J$ and $g$ being time-dependent, we can choose different ramping indices for them. As shown in Fig.~5a, the ramping trajectory in the parameter space of $J$ and $g$ depends on the ratio $r_{g}/r_{J}$, which affects the energy spectrum and the value of the energy gap. In Fig.~5c, we plot the energy spectrum of the lowest excited states vs the hopping rate $J$ for $r_{g}/r_{J}=1$, where the solid curve is the energy of the lowest symmetric state. The energy gap occurs at $J_{\rm gp}=0.104$ with $E_{\rm gp}=0.25$. The energy spectrum for $r_{g}/r_{J}\ne1$ can be found in Supplementary Figure~1a.
The sweeping rate of the Hamiltonian is $H_{\rm gp}^{\prime} = J^{\prime}_{\rm gp} I_{J} + g^{\prime}_{\rm gp}I_{g}$ with $I_{J}=\langle \partial H/\partial J \rangle_{\rm gp}$, $I_{g}=\langle \partial H/\partial g \rangle_{\rm gp}$, $g^{\prime}_{\rm gp}$ being the time derivative of the coupling $g$ at the gap position. With (\ref{eq:parat}) and (\ref{eq:path}), it can be shown that 
\begin{equation} 
g^{\prime}_{\rm gp}=\frac{r_{g}}{r_{J}}\frac{g_{\rm gp}}{J_{\rm gp}-J(0)} J^{\prime}_{\rm gp}. \label{eq:dgdtinp}
\end{equation} 
For a given ratio $r_{g}/r_{J}$, $g^{\prime}_{\rm gp}/J^{\prime}_{\rm gp}$ is a constant that does not depend on the specific value of $r_{J}$ or $r_{g}$. The dependence of $H_{\rm gp}^{\prime}$ on the ramping indices can hence be charactered by the dependence of $J^{\prime}_{\rm gp}$ on $r_{J}$, which is shown in Fig.~5b. For $r_{g}/r_{J}=1$, $J^{\prime}_{\rm gp}$ has a minimum at $r_{J}^{\rm (min)}=0.234$. 
In Fig.~5d, we plot the fidelity of the final state vs $r_{J}$ from our numerical simulation, which indicates that the best fidelity can be achieved when $r_{J}\in (1/2,1)$ at $T=10{\rm \pi}/g, 15{\rm \pi}/g$ and when $r_{J}\in (1/3,1/2)$ at $T=5{\rm \pi}/g$. This result confirms our analysis that the optimal ramping index for this trajectory will shift to a smaller value with $r_{J}^{\rm (min)}< 1$ in comparison to that of Fig.~3d. The discrepancy between the numerical and the estimated results of $r_{J}^{\rm (min)}$ could be owing to the small difference $\vert J_{\rm gp}-J(T)\vert$ between the gap position and the target parameter, which affects the accuracy of the Landau-Zener formula in adiabatic processes~\cite{lz1, lz2}. We also numerically simulate the ramping process for the target hopping rate $J(T)\in [0, 0.5]$ and obtain the fidelity of the final state vs $J(T)$ for several values of $r_{J}$, as plotted in Fig.~5e. The fidelity decreases as $J(T)$ becomes smaller, as $\vert J^{\prime}_{\rm gp}\vert$ increases with the difference $\vert J(T)-J(0)\vert$. 
The fidelity vs $r_{J}$ for $r_{g}/r_{J}\ne 1$ is given in Fig.~5f. It can be seen that the optimal ramping index $r_{J}^{\rm (min)}$ for different $r_{g}/r_{J}$ can be quite different. This is due to the change of the ramping trajectory and the energy spectrum as $r_{g}/r_{J}$ is varied. Detailed results on $E_{\rm gp}$, $J_{\rm gp}$, and $r_{J}^{\rm (min)}$ for different values of $r_{g}/r_{J}$ can be found in Supplementary Figure~1b.

\begin{figure}[t]
\centering
\includegraphics[width=8.5cm, clip]{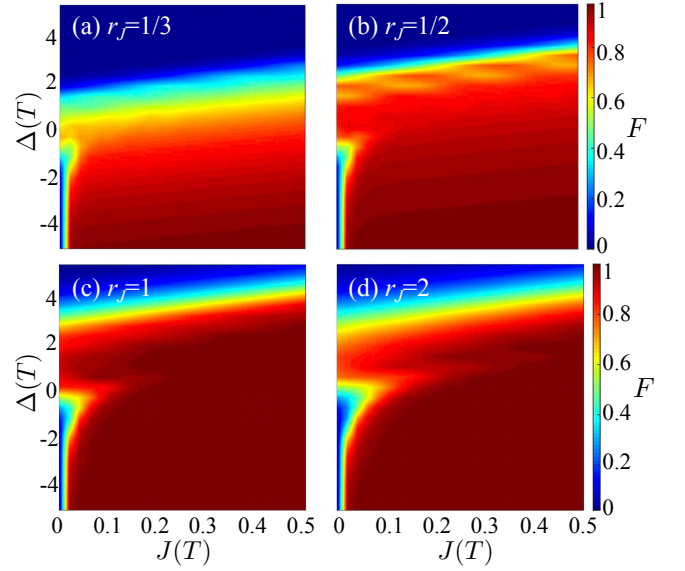}
\caption{{\bf Fidelity with SF initial state. a-d} Fidelity of the prepared state $F$ vs target hopping rate $J(T)$ and target detuning $\Delta(T)$ for ramping index $r_{J}$ given in the panel. Here $g(0)=0$, $g(T)=1$, $J(0)=0.5$, $\Delta(0)=0$, $r_{g}/r_{J}=r_{\Delta}/r_{J}=1$, and $T=15{\rm \pi}/g$.}
\label{fig6}
\end{figure}
We present the fidelity of the final state vs the target parameters following the trajectory (\ref{eq:parat}) with $g(0)=0$, $g(T)=1$, $J(0)=0.5$, $\Delta(0)=0$, $r_{g}/r_{J}=r_{\Delta}/r_{J}=1$, and the SF initial state (\ref{eq:Gg0}) for $r_{J}=1/3,\,1/2,\,1,\,2$, respectively, in Fig.~6a-d. Our numerical result shows that the fidelity decreases quickly as the target parameters enter the MI phase and strongly depends on the ramping index in the intermediate regimes of the parameter space.
\vskip 4mm

{\parindent 0 pt \bf DISCUSSION} 
\vskip 2mm

{\parindent 0 pt We have shown that the fidelity of the prepared state in the intermediate regimes of the parameter space can be improved by choosing the optimal ramping index for a given trajectory and by increasing the total ramping time $T$. Another approach to increase the fidelity is by choosing a favorable trajectory for a given set of target parameters. When the target parameters are in the MI phase, it is better to start from an initial state in the deep MI regime such as (\ref{eq:GJ0}) so that the adiabatic evolution does not need to cross a region with narrow energy gap to reach the target parameters so that diabatic transitions can be negligible. Similarly, when the target parameters are in the SF phase, we can choose the initial state to be in the deep SF regime such as (\ref{eq:Gg0}).  Combing the selection of the initial state with optimized nonlinear ramping can have dramatic impact on the fidelity of the prepared states. For illustration, in Fig.~7a, we plot the maximal fidelity among all eight sets of data in Fig.~4a-d and Fig.~6a-d for two initial states and various values of linear or nonlinear ramping index $r_{J}$. It can be seen that the maximal fidelity remains close to unity in almost the entire parameter space. For comparison, in Fig.~7b, we plot the maximal fidelity between the data for linear ramping ($r_{J}=1$) in Fig.~4c and Fig.~6c. The result in Fig.~7a outperforms that of Fig.~7b, and both results are much better than the individual plots in Fig.~4 and Fig.~6. We expect that further improvement can be achieved by optimizing the trajectory, e.g., using optimized $r_{g}/r_{J}, r_{\Delta}/r_{J}$ or the optimal control technique.}
\begin{figure}[t]
\centering
\includegraphics[width=8.5cm, clip]{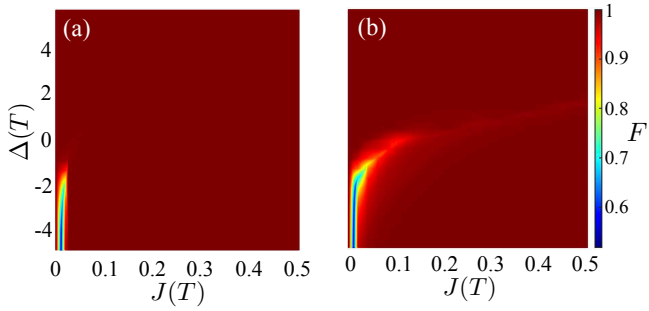}
\caption{{\bf Maximal fidelity with different ramping indices and trajectories.} Maximal fidelity of the prepared state $F$ vs target hopping rate $J(T)$ and target detuning $\Delta(T)$: {\bf a} among all data in Fig.~4a-d and Fig.~6a-d, and {\bf b} between the linear ramping data in Fig.~4c and Fig.~6c.}
\label{fig7}
\end{figure}

An obvious approach to improve the fidelity of adiabatic processes is to increase the ramping time $T$, which can reduce the time derivatives of the parameters and the sweeping rate of the Hamiltonian. This can be seen from the numerical result in Fig.~3d and Fig.~5d. For quantum devices in the NISQ era, however, the decoherence times of qubits and cavity modes set a limitation on the evolution time. The many-body ground states studied here involve finite number of polariton excitations. In the presence of decoherence, excitations can decay in a timescale comparable to the decoherence times. The ramping time needs to be much shorter than the decoherence times. In experiments, superconducting resonator cavities with frequency $\omega_{\rm c}/2{\rm \pi}=10$ GHz and quality factor $Q=10^{5}$ can be readily realized, which corresponds to a decay time of $1.6\,{\rm \mu s}$. Superconducting qubits can have a decoherence time of $\sim100\,{\rm \mu s}$~\cite{squbit_rev3}. With a typical coupling strength of $g/2\rm{\pi}, J/2\rm{\pi}=200$ MHz, the evolution time $T=15\rm{\pi}/g \approx 37.5$ ns. The state initialization pulses can be completed within a few $10$'s of ns. These time scales are much shorter than the decoherence times. To quantitatively characterize the effect of dissipation, we utilize a phenomenological approach with a non-Hermitian Hamiltonian~\cite{QOapproach}: $\widetilde{H}_{\rm t}= H_{\rm t} -{\rm i}(\kappa/2) \sum_{j} a_{j}^{\dag}a_{j} - {\rm i}(\gamma/2) \sum_{j} \sigma_{jz}$, where $\kappa$ is the cavity damping rate and $\gamma$ is the qubit decay rate. We numerically simulate the ramping process under the Hamiltonian $\widetilde{H}_{\rm t}$ and calculate the fidelity of the final state. For $r_{J}=1$ and $T=15{\rm \pi}/g$ studied in Fig.~3d, with $Q=5\times10^{4}$ ($\kappa/2{\rm \pi}=200$ kHz) and $\gamma/2{\rm \pi}=2$ kHz, the fidelity $F=0.9737$. For comparison, $F=0.9738$ when dissipation is not included. This result shows that the effect of dissipation with practical parameters is negligible at time scales of interest. Details of this result can be found in Supplementary Discussion and Supplementary Figure~2. Note that dissipation can be used for robust preparation and stabilization of entangled states, as studied in \cite{TureciPRA2014, TureciPRX2016}.
In addition, the qubit-cavity interaction in (\ref{eq:Hint}) has the form of JC coupling with the counter-rotating terms omitted. This is because we study the system in the strong-coupling regime with $g\ll \omega_{\rm c}, \omega_{z}$, where the effect of the counter-rotating terms can be neglected. 
\vskip 4mm

{\parindent 0 pt \bf METHODS} 
\vskip 2mm

{\parindent 0 pt We employ a method developed in our previous work \cite{Seo2015:1} to conduct numerical simulation on the JC lattice studied in this work. This method allows us to efficiently solve the ground states and the dynamics of a finite-sized JC lattice with given number of polariton excitations. In a JC lattice, each unit cell contains a qubit and a cavity mode. If we choose the photon cutoff on each lattice site to be equal to the total number of excitations in the lattice $N$, then the total number of basis states for this lattice is $(2N+1)^{L}$, which depends exponentially on the size of the lattice $L$. For $N=L=6$, the number of basis states is $4826809$. This dependence sets a serious limitation on the computable size of JC lattices. In our method, we only consider basis states that have exactly $N$ excitations. We developed a code to find out such basis states. For $N=L=6$, we find that the total number of basis states is $5336$, which shows that this method can greatly reduce the demand on computing power. We also developed codes to derive the matrix elements for the Hamiltonian and other operators, such as the creation and annihilation operators of cavity modes, the Pauli operators of qubits, and the number operator of cavity modes, under these basis states. With these matrices, we can calculate the eigenstates and eigenenergies of a JC lattice using the exact diagonalization method. We also simulate the dynamical evolution of this system under time-dependent Hamiltonians. The matrices for the creation operators and the Pauli operators enable us to generate the initial state in the MI and the SF phases numerically. Our method can be applied to different types of JC lattices as it can be used regardless of the specific form of interaction between neighboring lattice sites.}
\vskip 4mm

{\parindent 0 pt \bf DATA AVAILABILITY}
\vskip 2mm

{\parindent 0 pt The data that support the findings of this study are available from the authors upon reasonable request.}
\vskip 4mm

{\parindent 0 pt \bf CODE AVAILABILITY}
\vskip 2mm

{\parindent 0 pt The codes that are used to produce the data presented in this study are available from the authors upon reasonable request.}
\vskip 4mm

{\parindent 0 pt \bf ACKNOWLEDGEMENTS} 
\vskip 2mm

{\parindent 0 pt This work is supported by the UC Multicampus-National Lab Collaborative Research and Training under Award No. LFR-17-477237. G.L.L. is supported by the National Natural Science Foundation of China under Grant No. 11974205 and No. 11774197. C.W.W. is also supported by NSF QII-TAQS-1936375, NSF 1919355, and ONR N00014-15-1-2368. L.T. is also supported by UC Merced Faculty Research Grants 2017 and NSF awards No.1720501, No. 2006076, and No. 2037987.}
\vskip 4mm

{\parindent 0 pt \bf AUTHOR CONTRIBUTIONS}
\vskip 2mm

{\parindent 0 pt K.C. and P.P. are co-first authors. K.C. and P.P. conducted numerical simulation, L.T. designed the project and conducted analytical derivation, K.C., P.P. and L.T. analyzed the numerical data and wrote the paper with inputs from G.-L.L. and C.W.W., all authors discussed the results and contributed to the final paper.}
\vskip 4mm

{\parindent 0 pt \bf COMPETING INTERESTS}
\vskip 2mm

{\parindent 0 pt The authors declare that there are no competing interests.}
\vskip 4mm

{\parindent 0 pt \bf ADDITIONAL INFORMATION}
\vskip 2mm
{\parindent 0 pt {\bf Supplementary information} The online version contains supplementary material available at:}

\vskip 4mm

{\parindent 0 pt \bf FIGURE LEGENDS} 
\vskip 2mm

{\parindent 0 pt {\bf Fig.~1 Quantum phase transition in JC lattice. a} Schematic of a 1D JC lattice. Circles (rectangles) represent qubits (cavity modes) with light-matter coupling $g$ and hopping rate $J$. {\bf b} Single-particle density matrix $\rho_{1}(1,4)$ vs hopping rate $J$ and detuning $\Delta$ for a finite-sized lattice at unit filling with $N=L=6$. Here we let $g$ be the energy unit with $g\equiv1$.}
\vskip 2mm

{\parindent 0 pt {\bf  Fig.~2 Pulse sequence for state initialization. a} Pulses for MI initial state at $J=0$ and finite $g$. The vertical arrows are Rabi flips between the states $\vert g_{0}\rangle$ and $\vert 1,-\rangle$ in each JC model. {\bf b} Pulses for SF initial state at $g=0$ and finite $J$. The vertical (slanted) arrows are the operations $C_{l}$ ($Q_{l}$) with $l\in [1, N]$ on the coupled system of the auxiliary qubit and mode $a_{k=0}$.}
\vskip 2mm

{\parindent 0 pt {\bf Fig.~3 Ramping from MI to SF phase. a} Energy spectrum of the lowest excited states vs hopping rate $J$. Solid (dotted) curve is for the symmetric (asymmetric) state with the ground-state energy set to zero. 
{\bf b} Time derivative $J^{\prime}_{\rm gp}$ vs ramping index $r_{J}$ at $J(T)=0.5$ and $T=5{\rm \pi}/g, 10{\rm \pi}/g, 15{\rm \pi}/g$ from top to bottom.
{\bf c} Fidelity $F$ vs $J(T)=J$ for $r_{J}=2$ (solid), $1$ (dashed), $1/2$ (dot-dashed), and $1/3$ (dotted) at $T=15{\rm \pi}/g$.
{\bf d} Fidelity $F$ vs $r_{J}$ at $J(T)=0.5$ and $T=5{\rm \pi}/g, 10{\rm \pi}/g, 15{\rm \pi}/g$ from bottom to top.
In all plots, $g(t)\equiv1$, $J(0)=0$, and $\Delta(t)\equiv0$.}
\vskip 2mm

{\parindent 0 pt {\bf Fig.~4 Fidelity with MI initial state. a-d} Fidelity of the prepared state $F$ vs target hopping rate $J(T)$ and target detuning $\Delta(T)$ for ramping index $r_{J}$ given in the panel. Here $g(t)\equiv 1$, $J(0)=\Delta(0)=0$, $r_{\Delta}/r_{J}=1$, and $T=15{\rm \pi}/g$.}
\vskip 2mm

{\parindent 0 pt {\bf Fig.~5 Ramping from SF to MI phase. a} Ramping trajectory in the parameter space of $J$ and $g$ for $r_{g}/r_{J}=1/3, 1/2, 1, 2, 3$ from top to bottom.
{\bf b} Time derivative $J^{\prime}_{\rm gp}$ vs ramping index $r_{J}$ at $r_{g}/r_{J}=1$, $J(T)=0$, and $T=5{\rm \pi}/g, 10{\rm \pi}/g, 15{\rm \pi}/g$ from top to bottom.
{\bf c} Energy spectrum of the lowest excited states vs hopping rate $J$ for $r_{g}/r_{J}=1$. Solid (dotted) curve is for the symmetric (asymmetric) state with the ground-state energy set to zero. 
{\bf d} Fidelity $F$ vs $r_{J}$ at $r_{g}/r_{J}=1$, $J(T)=0$, and $T=5{\rm \pi}/g, 10{\rm \pi}/g, 15{\rm \pi}/g$ from bottom to top.
{\bf e} Fidelity $F$ vs $J(T)=J$ for $r_{J}=2$ (solid), $1$ (dashed), $1/2$ (dot-dashed), and $1/3$ (dotted) at $r_{g}/r_{J}=1$ and $T=15{\rm \pi}/g$.
{\bf f} Fidelity $F$ vs $r_{J}$ at $r_{g}/r_{J}=1/3$ (square), $1/2$ (circle), $1$ (diamond), $2$ (triangle), and $3$ (inverted triangle), $J(T)=0$, and $T=15{\rm \pi}/g$. 
In all plots, $g(0)=0$, $g(T)=1$, $J(0)=0.5$, and $\Delta(t)\equiv0$.}
\vskip 2mm

{\parindent 0 pt {\bf Fig.~6. Fidelity with SF initial state. a-d} Fidelity of the prepared state $F$ vs target hopping rate $J(T)$ and target detuning $\Delta(T)$ for ramping index $r_{J}$ given in the panel. Here $g(0)=0$, $g(T)=1$, $J(0)=0.5$, $\Delta(0)=0$, $r_{g}/r_{J}=r_{\Delta}/r_{J}=1$, and $T=15{\rm \pi}/g$.}
\vskip 2mm

{\parindent 0 pt {\bf Fig.~7. Maximal fidelity with different ramping indices and trajectories.} Maximal fidelity of the prepared state $F$ vs target hopping rate $J(T)$ and target detuning $\Delta(T)$: {\bf a} among all data in Fig.~4a-d and Fig.~6a-d, and {\bf b} between the linear ramping data in Fig.~4c and Fig.~6c.}


\begin{thebibliography}{99}
\bibitem{JCmodel}Jaynes, E. T. \& Cummings, F. W. Comparison of quantum and semiclassical radiation theories with application to the beam maser. \textit{Proceedings of the IEEE} \textbf{51}, 89-109 (1963).

\bibitem{cavityQED}Raimond, J. M., Brune, M. \& Haroche, S. Colloquium: Manipulating quantum entanglement with atoms and photons in a cavity. \textit{Rev. Mod. Phys.} \textbf{73}, 565-582 (2001).

\bibitem{circuitQED1}Girvin, S. M. Basic concepts in quantum information. \textit{Strong Light-Matter Coupling: from Atoms to Solid-State Systems} 155-206 (World Scientific, Singapore, 2013).

\bibitem{circuitQED2}You J. Q. \& Nori, F. Atomic physics and quantum optics using superconducting circuits. \textit{Nature} \textbf{474}, 589-597 (2011).

\bibitem{Hartmann:2006}Hartmann, M. J., Brand\~{a}o, F. G. S. L. \& Plenio, M. B. Strongly interacting polaritons in coupled arrays of cavities. \textit{Nat. Phys.} \textbf{2}, 849-855 (2006).

\bibitem{Greentree:2006}Greentree, A. D., Tahan, C., Cole, J. H. \& Hollenberg, L. C. L. Quantum phase transitions of light. \textit{Nat. Phys.} \textbf{2}, 856-861 (2006).

\bibitem{Angelakis:2007}Angelakis, D. G., Santos, M. F., \& Bose, S. Photon-blockade-induced Mott transitions and XY spin models in coupled cavity arrays. \textit{Phys. Rev. A} \textbf{76}, 031805(R) (2007).

\bibitem{2007RossiniPRL_JC}Rossini, D. \& Fazio, R. Mott-insulating and glassy phases of polaritons in 1D arrays of coupled cavities. \textit{Phys. Rev. Lett.} {\bf 99}, 186401 (2007).

\bibitem{2008NeilPra_BH}Na, N., Utsunomiya, S., Tian, L. \& Yamamoto, Y. Strongly correlated polaritons in a two-dimensional array of photonic crystal microcavities. \textit{Phys. Rev. A} {\bf 77}, 031803(R) (2008).

\bibitem{2009KochPra_QS}Koch, J. \& Le Hur, K. Superfluid-Mott-insulator transition of light in the Jaynes-Cummings lattice. \textit{Phys. Rev. A} {\bf 80}, 023811 (2009).

\bibitem{2012HouckNP_JCQS} Houck, A. A., T\"{u}reci, H. E. \& Koch, J. On-chip quantum simulation with superconducting circuits. \textit{Nat. Phys.} {\bf 8}, 292-299 (2012).

\bibitem{TianPRL2011}Hu, Y. \& Tian, L. Deterministic generation of entangled photons in superconducting resonator arrays. \textit{Phys. Rev. Lett.}  \textbf{106}, 257002 (2011).

\bibitem{Seo2015:1}Seo, K. \& Tian, L. Quantum phase transition in a multiconnected superconducting Jaynes-Cummings lattice. \textit{Phys. Rev. B}  \textbf{91}, 195439 (2015).

\bibitem{2015TianScienceChina_QS}Seo, K. \& Tian, L. Mott insulator-superfluid phase transition in a detuned multi-connected Jaynes-Cummings lattice.  \textit{Sci. China-Phys. Mech. Astron.} \textbf{58}, 070302 (2015).

\bibitem{Xue2017}Xue, J., Seo, K., Tian, L. \& Xiang, T. Quantum phase transition in a multiconnected Jaynes-Cummings lattice. \textit{Phys. Rev. B} \textbf{96}, 174502 (2017).

\bibitem{Hoffman:2011}Hoffman, A. J. et al. Dispersive photon blockade in a superconducting circuit. \textit{Phys. Rev. Lett.} \textbf{107}, 053602 (2011).

\bibitem{KeelingPRL2012}Nissen, F. et al Nonequilibrium dynamics of coupled qubit-cavity arrays. \textit{Phys. Rev. Lett.} \textbf{108}, 233603 (2012).

\bibitem{HouckPRX2017}Fitzpatrick, M., Sundaresan, N. M., Li, A. C. Y., Koch, J. \& Houck, A. A. Observation of a dissipative phase transition in a one-dimensional circuit QED lattice. \textit{Phys. Rev. X} \textbf{7}, 011016 (2017).

\bibitem{TureciPRL2012}Schir\'{o}, M., Bordyuh, M., \"{O}ztop, B. \& T\"{u}reci, H. E. Phase transition of light in cavity QED lattices. \textit{Phys. Rev. Lett.} \textbf{109}, 053601 (2012).

\bibitem{JalalPRA2013}Kumar, B. \& Jalal, S. Quantum Ising dynamics and Majorana-like edge modes in the Rabi lattice model. \textit{Phys. Rev. A} \textbf{88}, 011802(R) (2013).

\bibitem{Feynman}Feynman, R. P. Simulating physics with computers. \textit{Int. J. Theor. Phys.} \textbf{21}, 467-488 (1982).  

\bibitem{Lloyd}Lloyd, S. Universal quantum simulators. \textit{Science} \textbf{273}, 1073-1078 (1996).

\bibitem{AspuruGuzik2005}Aspuru-Guzik, A., Dutoi, A. D., Love, P. J. \& Head-Gordon, M. Simulated quantum computation of molecular energies. \textit{Science} \textbf{309}, 1704-1707 (2005).

\bibitem{Farhi2000_1}Farhi, E., Goldstone, J., Gutmann, S. \& Sipser, M. Quantum computation by adiabatic evolution. Preprint at http://
arxiv.org/abs/quant-ph/0001106 (2000).


\bibitem{Albash2016}Albash T. \& Lidar, D. A. Adiabatic quantum computation. \textit{Rev. Mod. Phys.} \textbf{90}, 015002 (2018).

\bibitem{FarhiScience2001}Farhi, E. et al. A quantum adiabatic evolution algorithm applied to random instances of an NP-complete problem. \textit{Science} \textbf{292}, 472-476 (2001).

\bibitem{RolandCerfPRA2002}Roland, J. \& Cerf, N. J. Quantum search by local adiabatic evolution. \textit{Phys. Rev. A} \textbf{65}, 042308 (2002).

\bibitem{HTQuanNJP2010}Quan H. T. \& Zurek, W. H. Testing quantum adiabaticity with quench echo. \textit{New J. Phys.} \textbf{12}, 093025 (2010).

\bibitem{XChenPRL2010}Chen, X., Lizuain, I., Ruschhaupt, A., Gu\'{e}ry-Odelin, D. \& Muga, J. G. Shortcut to adiabatic passage in two- and three-level atoms. \textit{Phys. Rev. Lett.} \textbf{105}, 123003 (2010).

\bibitem{delCampoPRL2012}del Campo, A., Rams, M. M. \& Zurek, W. H. Assisted finite-rate adiabatic passage across a quantum critical point: exact solution for the quantum Ising model. \textit{Phys. Rev. Lett.} \textbf{109}, 115703 (2012).

\bibitem{Kitaev1995}Kitaev, A. Yu. Quantum measurements and the Abelian stabilizer problem. Preprint at http://arxiv.org/abs/quant-ph/9511026 (1995).

\bibitem{Abrams1997} Abrams, D. S. \& Lloyd, S. Simulation of many-body Fermi systems on a universal quantum computer. \textit{Phys. Rev. Lett.}   \textbf{79}, 2586-2589 (1997).

\bibitem{Peruzzo2014}Peruzzo, A. et al. A variational eigenvalue solver on a photonic quantum processor. \textit{Nat. Commun.} \textbf{5}, 4213 (2014).

\bibitem{Dumitrescu2018}Dumitrescu, E. F.  et al. Cloud quantum computing of an atomic nucleus. \textit{Phys. Rev. Lett.} \textbf{120}, 210501 (2018).

\bibitem{LongResearch2020}Wei, S. J., Li, H \& Long, G. L. A full quantum eigensolver for quantum chemistry simulations. \textit{Research} \textbf{2020}, 1486935 (2020).

\bibitem{Kraus2008} Kraus, B. et al. Preparation of entangled states by quantum Markov processes. \textit{Phys. Rev. A} \textbf{78}, 042307 (2008).

\bibitem{Verstraete2009} Verstraete, F., Wolf, M. M. \& Cirac, J. I. Quantum computation and quantum-state engineering driven by dissipation. \textit{Nat. Phys.} \textbf{5}, 633-636 (2009).

\bibitem{TureciPRA2014} Aron, C., Kulkarni, M. \& T\"{u}reci, H. E. Steady-state entanglement of spatially separated qubits via quantum bath engineering. \textit{Phys. Rev. A} \textbf{90}, 062305 (2014).

\bibitem{TureciPRX2016} Aron, C., Kulkarni, M. \& T\"{u}reci, H. E. Photon-mediated interactions: a scalable tool to create and sustain entangled states of N atoms. \textit{Phys. Rev. X} \textbf{6}, 011032 (2016).

\bibitem{Preskill}Preskill, J. Quantum computing in the NISQ era and beyond. \textit{Quantum} \textbf{2}, 79 (2018).

\bibitem{SenPRL2008}Sen, D., Sengupta, K. \& Mondal, S. Defect production in nonlinear quench across a quantum critical point. \textit{Phys. Rev. Lett.}  \textbf{101}, 016806 (2008).

\bibitem{MondalPRB2009}Mondal, S., Sengupta, K. \& Sen, D. Theory of defect production in nonlinear quench across a quantum critical point.  \textit{Phys. Rev. B} \textbf{79}, 045128 (2009).

\bibitem{BarankovPRL2008}Barankov, R. and Polkovnikov, A. Optimal nonlinear passage through a quantum critical point. \textit{Phys. Rev. Lett.}  \textbf{101}, 076801 (2008). 

\bibitem{lz1}Landau, L. D. Zur theorie der energieubertragung ii. \textit{Phys. Z. Sowjetunion} \textbf{2}, 46-51 (1932).

\bibitem{lz2}Zener, C. Non-adiabatic crossing of energy levels. \textit{Proc. R. Soc. London. A} \textbf{137}, 696-702 (1932). 

\bibitem{1996EberlyPRL_StatePrepare}Law, C. K. \& Eberly, J. H. Arbitrary control of a quantum electromagnetic field. \textit{Phys. Rev. Lett.} {\bf 76}, 1055-1058 (1996).

\bibitem{NeillScience2018}Neill, C. et al. A blueprint for demonstrating quantum supremacy with superconducting qubits. \textit{Science} \textbf{360},195-199 (2018). 

\bibitem{YYu2018}Zhao, P. et al. Two-photon driven Kerr resonator for quantum annealing with three-dimensional circuit QED. \textit{Phys. Rev. Applied}  \textbf{10}, 024019 (2018).

\bibitem{squbit_rev1}Devoret, M. H. \& Schoelkopf, R. J. Superconducting circuits for quantum information: an outlook.  \textit{Science} \textbf{339}, 1169-1174 (2013). 

\bibitem{squbit_rev2}Wendin, G. Quantum information processing with superconducting circuits: a review. \textit{Rep. Prog. Phys.} \textbf{80} 106001 (2017). 

\bibitem{squbit_rev3}Krantz, P. et al. A quantum engineer's guide to superconducting qubits. \textit{Appl. Phys. Rev.} \textbf{6}, 021318 (2019).  

\bibitem{TianPRB2013}Mei, F., Stojanovi\'{c}, V. M., Siddiqi, I. \& Tian, L. \textit{An analog superconducting quantum simulator for Holstein polarons}, Phys. Rev. B \textbf{88}, 224502 (2013).

\bibitem{Penrose:1956}Penrose, O. \& Onsager, L. Bose-Einstein condensation and liquid helium. \textit{Phys. Rev.} \textbf{104}, 576-584 (1956).

\bibitem{Yang:1962}Yang, C. N. Concept of off-diagonal long-range order and the quantum phases of liquid He and of superconductors. \textit{Rev. Mod. Phys.} \textbf{34}, 694-704 (1962).


\bibitem{QOapproach}Dalibard, J., Castin, Y. \& M\o lmer, K. Wave-function approach to dissipative processes in quantum optics. \textit{Phys. Rev. Lett.} \textbf{68}, 580-583 (1992).

\end{thebibliography}
\end{document}